\documentclass[]{aa}
\bibstyle{aa}
\usepackage{graphicx}

\usepackage{aas_macros}

\usepackage{natbib}

\usepackage{float}
\makeatletter

\newcommand{\Rmnum}[1]{\expandafter\@slowromancap\romannumeral #1@}

\begin{document}
\title{Turbulent entrainment origin of  protostellar outflows}
 \author{Guang-Xing Li \inst{1} \and Keping Qiu \inst{1,2} \and Friedrich Wyrowski\inst{1} \and Karl Menten \inst{1}}
 \institute{Max-Planck Institut f\"ur Radioastronomie, Auf dem H\"ugel, 69,  53121 Bonn, Germany \and School of Astronomy and Space Science, Nanjing University, Nanjing 210093, China}
\offprints{Guang-Xing Li, \email{gxli@mpifr-bonn.mpg.de}}
\titlerunning{Turbulent outflow model}

\abstract{ Protostellar outflow is a prominent process that accompanies the
formation of stars. It is generally agreed that wide-angled protostellar
outflows come from the interaction between the wind from a forming  star and the
ambient gas. However, it is still unclear how the interaction takes place.
In this work, we theoretically investigate the possibility that the outflow
results from interaction between the wind and the ambient gas in the form of
turbulent entrainment. In contrast to the previous models, turbulent motion of
the ambient gas around the protostar is taken into account.
In our model, the ram-pressure of the wind balances the turbulent ram-pressure
of the ambient gas, and the outflow consists of the ambient gas entrained by the
wind.
The calculated outflow from our modelling exhibits a conical shape. The total
mass of the outflow is determined by the turbulent velocity of the envelope as
well as the outflow age, and the velocity of the outflow is several times higher
than the velocity dispersion of the ambient gas. The outflow opening angle
increases with the strength of the wind and decreases with the increasing
ambient gas turbulence. The outflow exhibits a broad line width at every
position. We propose that the turbulent entrainment process, which happens
ubiquitously in nature, plays a universal role in shaping protostellar outflows.
}


\maketitle

\section{Introduction}
Protostellar outflow is a prominent process that is intimately connected with
the formation of stars. When a protostar accretes gas from its surroundings, it
produces a powerful wind or jet. As the wind or jet material moves away from the
protostar, ambient gas is entrained. The mixture of the wind or jet gas and the
ambient gas will move away from the protostar, forming a so-called molecular
outflow. In observations of the star-forming regions with rotational transitions
of molecules such as CO, outflows can be easily identified from the
high-velocity part of the emission line. The physical size of the molecular
outflow is about a parsec. The typical  velocity of the outflowing gas is about
tens of kilometers per second, and the morphology of the outflow can be
collimated (jet-like) or less collimated (conical-shaped), or a combination of
both.

Many of the outflows exhibit as conical-shaped geometry
\citep{1998ApJ...507..861S,2000ApJ...542..925L,2006ApJ...646.1070A,2009ApJ...696...66Q,2011MNRAS.415L..49R,2011ApJ...729..124C}.
 It is still unclear how these conical-shaped outflows form.
Models that can explain conical-shaped outflows include the wind-driven shell
model \citep{1991ApJ...370L..31S,1996ApJ...472..211L,2001ApJ...557..429L}  and
the circulation model \citep{1996MNRAS.281.1038F,1999A&A...350..254L}. In the
wind-driven shell model, the wind from the protostar blows into the ambient
medium, and a shell forms at the interface between the wind and the ambient gas.
 The shell absorbs momentum from the wind and expands. The  outflowing
material consists of the ambient gas swept up by the shell. In the circulation
model, the gas circulates around the protostar in a quadrupolar way, and it is
the gas that moves outward in this circulation cycle that constitutes the
outflow.

Neither of these models properly addresses the role of supersonic turbulent
motion in the formation and evolution of the protostellar outflows. This might be
because the importance of supersonic turbulent motion in star-forming regions
was not generally recognized when these models were proposed. Today, it is
widely recognized that the interstellar medium where stars form is turbulent,
 this turbulence is supersonic in most star-forming regions
\citep{1981MNRAS.194..809L,2007prpl.conf...63B}. This will bring about serious
physical discrepancies to both models. For the wind-driven shell model, the
turbulent ram-pressure (expressed as $p_{\rm turb}\sim \rho \sigma_{\rm
turb}^2$) is much higher than the thermal pressure, and is often comparable to
the ram-pressure of the wind (expressed as $p_{\rm wind}\sim\rho_{\rm
wind}v_{\rm wind}^2$, see section \S \ref{sec:discu}). Therefore the wind-driven
shell model is no longer valid since the physical condition cannot allow the
shell to expand for long. The circulation model is also questionable when the
interstellar medium is turbulent, because turbulent motion is expected to
significantly distort the motion of gas, making the initial condition
unfavourable for circulation to occur.

The kinematic structures of the outflows are only poorly reproduced by the
previous models. Mappings of outflows with molecular lines reveal the spatial and
velocity structures of the outflowing gas, which is crucial to test the models.
It is found that the outflows ubiquitously exhibit a broad line width at
different positions. Seen from position-velocity diagrams, the outflowing gas
exhibits a wide velocity spread at one given position (\S \ref{sec:obs}).
This is difficult to understand in the context of the widely accepted
wind-driven shell model
\citep{1991ApJ...370L..31S,1996ApJ...472..211L,2001ApJ...557..429L}, because the
shell swept by the wind will expand according to the { ``Hubble law''}, and
the expansion speed of the shell is higher in the directions where the shell has
progressed further. Seen from the position-velocity diagram , we can identify
two velocity components at one given position. One component comes from the
front wall of the outflow cavity, the other from the back wall of the outflow
cavity \citep{2000ApJ...542..925L}. However, observations found that at one single
position, there is only one broad component \citep[e.g.
][]{2009ApJ...696...66Q}.

It is observationally found that the opening angles of outflows evolve
significantly during their lifetimes
\citep{2005ccsf.conf..105B,2006ApJ...646.1070A}. While usually interpreted as
the evolution of the wind-envelope interaction \citep{2006ApJ...646.1070A}, a
comprehensive treatment of this interaction in the presence of the turbulent
motion of the ambient gas is still lacking.

In this study, we theoretically investigate the formation of outflows from the
turbulent mixing between the wind from the protostar and the {\it turbulent}
ambient gas. In our {\it wind-driven turbulent entrainment} model, the
ram-pressure of the wind and the turbulent ram-pressure of the ambient gas
(envelope) establish hydrostatic balance, and the outflow is the gas contained
in the turbulent entrainment layer between the wind and the turbulent envelope.
In \S \ref{sec:model} we present the basic physical picture, followed by a
detailed description of our model. In \S \ref{sec:obs} we compare the image and
position-velocity diagram from our model with existing observations. In \S
\ref{sec:collimation} and \ref{sec:massvelo} we offer a unified framework to
understand outflows from low-mass protostars and high-mass protostars as well as
AGN-driven outflows. We also discuss the possibility that the wind is not
strong enough to push the envelope, and predict the existence of dwarf
outflows in these situations (\S \ref{sec:drawf}). In \S \ref{sec:conclu} we
conclude.

Recently, the interaction between a wind and a turbulent core has been studied
with radiation-hydrodynamic simulations \citep{2011ApJ...743...91O}, focusing on
how the physical structure suggested by the simulations can be reproduced
through millimetre-wavelength aperture synthesis observations. In this work, we
present an explicit treatment of the mixing entrainment process, and study the
consequences of this turbulent entrainment in producing protostellar outflows.

\label{sec:physical}
\begin{figure} \includegraphics[width=0.35 \textwidth]{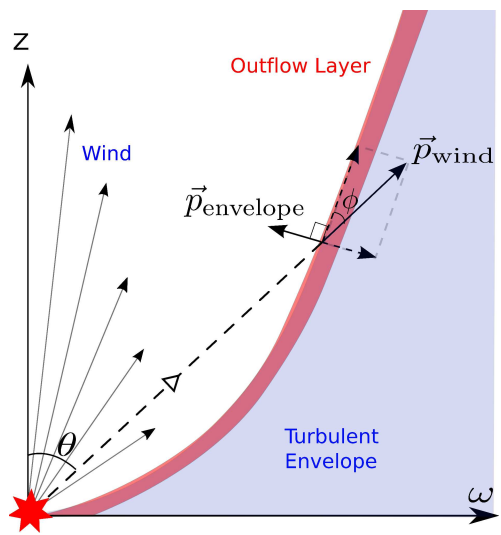}
\caption{Cartoon showing the basic structure of our model. The outflow (red region) lies between the wind and
the envelope.  Arrows ($\vec p_{\rm wind}$ and $\vec p_{\rm envelope}$) denote the ram-pressure of the wind and the envelope.
With {\bf wind} we denote the cavity evacuated by the wind, with {\bf turbulent
envelope} we denote the turbulent envelope, and  with {\bf outflow layer} we
denote the layer which contains the mixture of gas from the wind and gas from
the envelope. \label{balance} }
\end{figure}
\label{physical_picture}

\section{Model}\label{sec:model}
In our model, the outflowing gas is contained in the turbulent entrainment layer
that develops between the wind and the envelope (Fig. \ref{balance}). The wind
is collimated, and its strength reaches a maximum around the $z$-axis of Fig.
\ref{balance}. The wind is launched in a region close to the protostar due to
magnetic centrifugal forces. This wind is generally collimated and exhibits
universal asymptotic collimation properties
\citep{1995ApJ...455L.155S,1999ApJ...526L.109M}. The wind evacuates a cavity and
is confined mainly inside it. The ambient matter (envelope) is still present in
regions father away from the outflow axis. The outflow layer lies between the
cavity and the ambient matter. The outflow is generated by the interaction
between the wind and the envelope.

The turbulent nature of the interstellar medium has been recognized for years.
It has been suggested that it plays an important role in regulating the star
formation processes inside molecular clouds
\citep{1997ApJ...476..730P,2005ApJ...630..250K,2007prpl.conf...63B}. Molecular
outflows have been suggested as major drivers of this turbulent motion
\citep{2007ApJ...659.1394M,2007ApJ...662..395N}. The
role of turbulence in the formation and evolution of the molecular outflow
remains unexplored.
It is easy to show (\S \ref{sec:collimation}) that at physical scales of about a
parsec (which is the typical physical scale of the outflow), turbulence
dominates the dynamical properties of the medium. It is therefore necessary to
consider the roles of turbulence consistently.

In this study, we propose that the structure of the outflow is determined by the
following processes:
First, we propose that the wind and the envelope will establish hydrostatic
balance, and this balance will determine the shape of the outflow. According to
Fig. \ref{balance},  the wind has a ram-pressure that points away from the
protostar (see $\vec p_{\rm wind}$ in Fig. \ref{balance}, where $|\vec p_{\rm
wind}| \equiv p_{\rm wind}=\rho_{\rm wind} v_{\rm wind}^2$), and the envelope
has a ram-pressure that is perpendicular to the wall of the outflow cavity
($\vec p_{\rm envelope}$).  According to the balance of these two forces, the
shape of the outflow cavity can be numerically calculated (\S \ref{sec:shape}).

Second, we propose that a turbulent mixing layer develops between the wind
and the envelope. When the wind moves relative to the envelope, Kelvin-Helmholtz
instability occurs. When the instability saturates, a mixing layer establishes.
This mixing layer grows as it absorbs mass and momentum from the envelope and
the wind. The mass and momentum of the mixing layer is determined by mass and
momentum conservations (\S \ref{sec:mass},\ref{sec:conserve}).

Turbulence plays an important role in both processes. First, turbulence
influences the turbulent ram-pressure of the envelope,  and in turn influences
the pressure balance between the wind and the envelope. Therefore, turbulence
influences the shape of the outflow (\S \ref{sec:shape}). Second, turbulence
changes the mixing process that occurs between the wind and the envelope.
Therefore, it influences the mass and velocity of the outflow (\S
\ref{sec:mass},\ref{sec:conserve}).

The structure of the molecular cloud is inhomogeneous at a variety of scales.
On a parsec scale, millimeter/sub-millimeter studies suggest that the density distribution of
molecular gas around the protostar is consistent with being spherically
symmetric and can be described as a power-law $\rho \sim r ^{\gamma}$
\citep[e.g.][]{2010MNRAS.406..102K,2011ApJ...726...97L}.

Following these author we parametrized the density distribution of the envelope
as \begin{equation}\label{rho_r} \frac{\rho}{\rho_0} = \big( \frac{r}{r_0}\big
)^{- k_{\rho}}  \;.  \end{equation} By assuming hydrostatic equilibrium, Equation \ref{rho_r} leads to the distribution of
the turbulent velocity $\sigma_{\rm turb}$ that takes the form 
\citep{1996ApJ...469..194M,1999ApJ...522..313M,2003ApJ...585..850M}

\begin{equation}
 \frac{\sigma_{\rm turb}}{\sigma_0} = \big(\frac{r}{r_0}\big )^{1- k_{\rho}/2} \;.  
\end{equation} 

The power-law index $k_{\rho}$ in Equation \ref{rho_r} is set  according to the
observational studies by
\citet{2002ApJ...566..945B,2002ApJS..143..469M,2010MNRAS.406..102K,2011ApJ...726...97L}.
In our canonical model, consulting the results in \citet{2011ApJ...726...97L} we
set a density of $3.35 \times 10^{-19}\rm \ g \ cm^{-3}$ at $r=0.3\;\rm pc$ for
Equation \ref{rho_r}. The parameter values are summarized in Table
\ref{paramss}. These parameters are set to be typical of high-mass protostars
\citep[e.g.][]{2010MNRAS.406..102K,2011ApJ...726...97L}. Furthermore, since our
model is scale-free, the results can be scaled to different situations (\S
\ref{sec:selfsim}).

\begin{table} \label{table:1}
\begin{tabular}{p{0.08 \textwidth }p{0.18 \textwidth }p{0.15 \textwidth }}

\hline Parameter& Definition & Fiducial Value\\ \hline \hline
\multicolumn{2}{c}{Envelope Parameters} \\ \hline
 $k_{\rho}$       & Power-law
index for density distribution & 1.785\\ 
$k_{\sigma}$ & Power-law index for the distribution of the turbulent velocity & 0.1075\\
$r_{ 0}$   & Characteristic radius
to specify the values of physical quantities & 0.3 pc \\ 
$\rho_{ 0}$&
Density at the characteristic radius & $3.35\times 10^{-19}\rm \  g \
cm^{-3}$\\
 $\sigma_{ 0}$& Velocity dispersion at the characteristic
radius & $4.08\rm\  km\ s^{-1}$\\  \hline
\multicolumn{2}{c}{Wind Parameters} \\ \hline
$v_{\rm wind}$ & Velocity of the wind &$600 \;\rm km \;s^{-1}$\\
$\dot M_{\rm wind}$ & Mass-loss rate of the wind & 0.5, 1, 1.5  $\times 10^{-3}\; \rm M_{\odot}\; yr^{-1}$\\
\hline

  \multicolumn{2}{c}{Other parameters} \\ \hline
$\alpha$ &Entrainment Efficiency  &0.1 \\ 
$\beta$ & Wind Efficiency & 0.3 \\
$M_{*}$ & Stellar Mass & 10 $M_{\odot}$\\
$t$ & Age of outflow & $1\times 10^{4}\rm\; yr$\\

\hline
\end{tabular} \caption{Summary of model parameters. }\label{paramss} \end{table}

\subsection{Wind from the embedded protostar}
In our model, the outflow consists of the envelope gas entrained by the wind.
To specify the wind, we followed the formalism used by
\citet{2001ApJ...557..429L}.
In a spherical coordinate system $(r, \theta, \phi)$, \citet{1995ApJ...455L.155S}
have shown that the radial velocity of an X-wind is roughly constant, and
the density distribution takes the form 
\begin{equation} \label{xwind}
\rho_{\rm wind}\sim \frac{1}{r^2 \sin^2 \theta}\;.  \end{equation} 
Equation \ref{xwind} is shown to be valid not only for the X-wind model, but also for
more general force-free winds \citep{1997ApJ...486..291O,1999ApJ...526L.109M}. 
It has been shown that the terminal velocity of the wind $v_{\rm wind}$ is 
approximately the same at different angles \citep{1994ApJ...429..808N}. 
In our
modelling, the wind is parametrized as

\begin{eqnarray} \rho_{\rm wind} &=& \rho_{\rm norm}  \nonumber \frac{1}{r^2
(\sin^2 \theta+\epsilon)}\\ v_{\rm wind}   &=& v_t  \;,
\end{eqnarray}
where $\epsilon =0.01$ is added to avoid the singularity at the pole
($\theta=0$) \citep{2001ApJ...557..429L}.
$v_t$ is the terminal velocity of the wind,
 and $\rho_{\rm norm}$ is related to the wind mass-loss
rate (rate at which the wind carries matter from the protostar, $\dot M_{\rm
wind}$) by mass conservation,
\begin{equation}
\int_{0}^{\pi}2\pi r^2 \sin \theta \rho_{\rm wind}(r) v_{\rm wind}{\rm d} \theta = \dot M_{\rm wind}\;.
\end{equation} 
We assumed that the wind has a terminal velocity of 600 $\rm km\; s^{-1}$ 
\citep[similar to][]{2005ApJ...631.1010C}. The values for the mass-loss rate
were chosen to match the values found by \citet{2004ApJ...608..330B}. Parameters for the wind are listed in Table \ref{paramss}.

In principle, both the mass-loss rate of the wind and the density distribution
of the envelope should  evolve during the star formation process. However for
simplicity, we only considered an envelope with a fixed density distribution and
a wind with a fixed mass-loss rate.
In practice, the timescale for local collapse is proportional to $\rho^{-1/2}$.
Therefore, the central region where the protostars form evolves faster
than the outer envelope, the density
structure of the envelope evolves more slowly than the evolution
of the wind mass-loss rate during the lifetime of the outflow.

\begin{figure} \includegraphics[width=0.5 \textwidth]{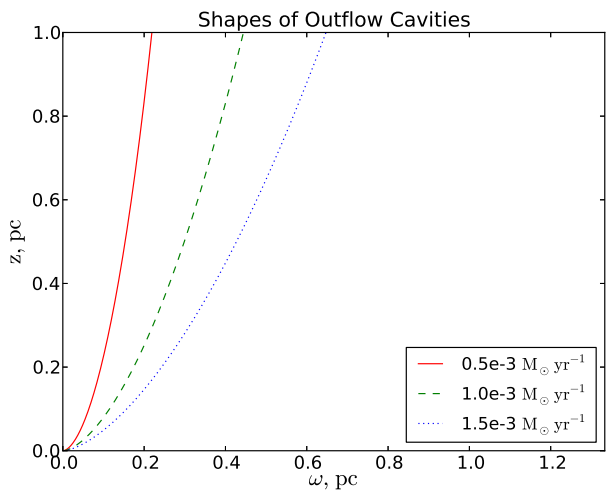}
\caption{Shapes of outflow cavities calculated for models with different wind
mass-loss rates. The parameters are taken from Table \ref{table:1}. See \S \ref{cavity_shape} for details.
\label{shapes}} \end{figure}
\hspace{0.3 cm}

\subsection{Shape of the outflow cavity \label{cavity_shape}\label{sec:shape}}

The shape of the outflow cavity is determined by the balance of the ram-pressure
of the wind and the pressure of the envelope perpendicular to the outflow cavity
wall (see Fig. \ref{balance}). The magnitude of $\vec p_{\rm wind}$ is $p_{\rm
wind}=\rho_{\rm wind} v_{\rm wind}^2$ , and the magnitude of the ram-pressure
vector  $\vec p_{\rm envelope}$ is $p_{\rm envelope}=\rho_{\rm
envelope}\sigma_{\rm turb}^2$.
Projected perpendicular to the edge of the outflow cavity, the magnitude of the
 wind pressure is $|\vec p_{\rm wind}| \sin \phi$ and  the magnitude of
the envelope pressure is $|\vec p_{\rm envelope}|$  (Figure \ref{balance}).
The {\it pressure balance } perpendicular to the wall of the outflow cavity can
then be written as

\begin{equation} \label{pbal}
p_{\rm wind}\sin\phi=p_{\rm envelope}\;.
\end{equation}
 Along the cavity wall, no pressure balance is established, and the wind
 contributes its momentum to the outflow (see \S\ \ref{sec:mass}). 

Here we used cylindrical coordinates
$(\omega,\phi,z )$ for convenience. The polar angle $\theta$ is
specified as (Figure \ref{balance})
 \begin{equation}\label{geom1}
\theta=\frac{\pi}{2}-\arctan\big(\frac{z}{\omega}\big)\;, 
\end{equation} 

and
the angle between the cavity wall and the $z$ axis is specified as

\begin{equation}\label{geom2} 
\theta_{\rm
wall}=\frac{\pi}{2}-\arctan\big(\frac{\rm d z}{\rm d \omega}\big)\;.
\end{equation} 

This gives a value $\phi$ of (Fig. \ref{balance})

\begin{equation}\label{eqphi}
\phi = \theta-\theta_{\rm wall}=\arctan\big(\frac{\rm d z}{\rm d \omega}\big)-\arctan\big(\frac{z}{\omega}\big)\;.
\end{equation}.

  Combining Equations \ref{pbal}, \ref{geom1}, \ref{geom2}, and \ref{eqphi}
  gives

\begin{equation}\label{master_balance} \frac{\rm d z}{\rm d
\omega}=\tan \phi=\tan \Big(\arctan \Big(\frac{z}{\omega}\Big)+ \arcsin \frac{p_{\rm envelope}}{p_{\rm
wind}}\Big) \;.\end{equation}

Equation \ref{master_balance}
determines the shape of the outflow cavity. It is a first-order differential
equation and can be solved numerically. We imposed the
boundary condition $\omega= 4.5 \times 10^{15}\rm \ cm$ at $z=0$, which is
similar to those used by \citet{2009MNRAS.400..629P}. The results are not
sensitive to the choice of this boundary condition. 

{ In deriving Equation \ref{master_balance}, we have neglected the effect of
centrifugal force. As will be shown in section \ref{sec:cf}, centrifugal force
is not significant in our case.}

Using Equation \ref{master_balance}, we calculated the shapes of the outflow
cavities for winds of different mass-loss rates. The results are shown in Fig.
\ref{shapes}. It is clear that the opening angle of the outflow depends on the
strength of the wind: a stronger wind leads to a less collimated outflow.

Observationally, the opening angle of the outflow evolves during the evolution
of the protostar \citep{2006ApJ...646.1070A,2005ccsf.conf..105B}. This is
interpreted as the evolution of the wind-envelope interaction
\citep{2006ApJ...646.1070A}. However, according to the widely accepted
wind-driven shell model, the opening angle of the outflow should be fixed as the
wind collimation is fixed and the density structure of the envelope does not
evolve much \citep{2001ApJ...557..429L}. How the evolution of  the wind-envelope
interaction changes the collimation of the outflow is thus unclear.

According to our model, the outflow opening angle is determined by the force
balance between the wind and the envelope. With the wind strength increases 
or the envelope becomes less turbulent, the opening angle of the outflow becomes
larger, and the outflow tends to decollimate. Our model provides a quantitative
explanation of the de-collimation of the outflow in the context of the
wind-envelope interaction.

Turbulence in the envelope plays a vital role in our model: the shape of the
outflow cavity is determined by the hydrostatic balance between the wind and the
envelope, and an increase in the strength of the wind can decrease the
collimation of the outflow, while an increase of the turbulence in the envelope
can increase the collimation of the outflow. Turbulence helps to {\it collimate}
the outflow.

\subsection{Mass entrainment of the outflow layer} \label{sec:mass}
\label{entrain_para}

As discussed in \S \ref{physical_picture}, the growth of the outflow depends on
the way matter is entrained into it.

The key parameter that characterizes the mass entrainment process is the local
mass entrainment rate, which is the amount of matter entrained per given time
interval per surface area.
The entrainment rate $\dot m$ can therefore be defined as
\begin{equation}
\dot m=\frac{M_{\rm entrained}}{{\rm d }s\times {\rm d} t}\;,
\end{equation}
where $M_{\rm entrained}$ is the amount of gas entrained, d$\rm s$ represents
the unit surface area and d$\rm t$ represents the time interval.
The entrainment rate depend on the density and the velocity involved in the entrainment process.
Therefore, the general form of the entrainment rate can be formulated as 
\begin{equation}
\dot m =  f_{\rm entrainment}(c_s, \sigma_{\rm turb},v_{\rm shear}) \times \rho_{\rm envelope}\;,
\end{equation}
where $c_s$, $\sigma_{\rm turb}$ and $v_{\rm shear}$ are the sound speed,
turbulent speed and the velocity difference between the shearing layer and the
envelope, and $\rho_{\rm envelope }$ is the density of the envelope.

The factor $f_{\rm entrainment}$ takes into account the effect of different
speeds on the entrainment rate, and has the dimension of the velocity.  In our
case, the relative speed between the outflow and the envelope $v_{\rm shear}$ is much
higher than the velocity dispersion of the envelope $\sigma_{\rm turb}$, and
$\sigma_{\rm turb}$ is higher than the sound speed ($v_{\rm shear}>>
\sigma_{\rm turb} >c_s$).

The entrainment process is driven by the shearing motion and limited by the
transport properties of the medium. The shearing motion is characterized by
$v_{\rm shear}$ and the transport process is characterized by $\sigma_{\rm
turb}$ and $c_{\rm s}$. As $v_{\rm shear}>> \sigma_{\rm turb}$, the entrainment
rate is limited by the transport and is determined by a combination of
$\sigma_{\rm turb}$ and $c_{\rm s}$. In a turbulent medium, turbulence
dominates the transport. Therefore, in our model, $f_{\rm entrainment}$ is
determined by turbulence. We thus take
\begin{equation}
f_{\rm entrainment}= 
\alpha \sigma_{\rm turb}\;,
\end{equation}
where $\alpha$ is a dimensionless factor that characterizes the efficiency of the entrainment process.
 We therefore parametrized the 
entrainment rate in the presence of turbulent motion as
\begin{equation}\label{entrain_turb} 
\dot m =\alpha \sigma_{\rm turb} \rho_{\rm envelope}\;,
\end{equation} 
where $\sigma_{\rm turb}$ is the characteristic velocity of turbulent motion in the medium. We assume $\alpha=0.1$ in this work.

Historically, turbulent mixing has long been proposed as a mean of entraining
mass into the outflow \citep[e.g.][]{1991ApJ...372..646C,2004ApJ...608..274W}.
However, these models are doubted because the entrainment rate is far
too small to account for the actual mass the outflow
\citep{1997ApJ...479L..59C}. Realizing the importance of
turbulence for enhancing the mass and momentum transport, we here propose that
the argument by \citet{1997ApJ...479L..59C} no longer holds in our cases because
turbulent mixing can be significantly enhanced when the ambient medium is
turbulent.

\subsection{Mass and momentum conservation in the entrainment
layer}\label{sec:conserve}
\begin{figure} \includegraphics[width=0.45 \textwidth]{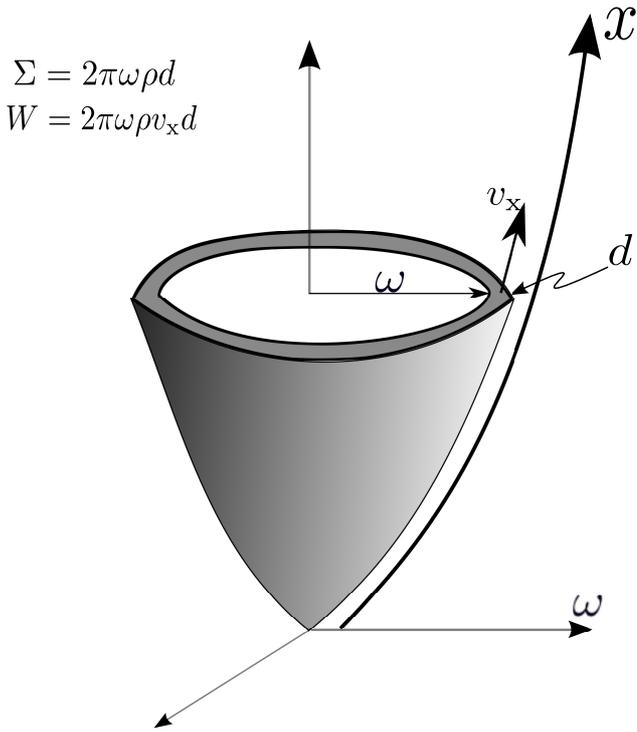}
\caption{Cartoon illustrating the geometry of the outflow layer and the
definition of physical quantities in \S \ref{sec:conserve} \label{entrain3d} .} \end{figure}

Based on the assumption that the outflow gas is the entrained envelope material,
we formulate the equations that govern the evolution of the outflow. The
geometry of the outflow is illustrated in Fig. \ref{entrain3d}.
We defined the $x$-axis to be along the wall of the outflow cavity and $v_{\rm
x}$ as the average velocity of the outflow along the x-axis.
By doing this, our $x$-axis is no longer a straight line. We used this
construction for simplicity.
 $\omega$ is
the same as illustrated in Fig. \ref{balance}, which gives a circumference
$l_c$ of 

\begin{equation} 
l_{\rm c}=2\pi \omega\;.
\end{equation} 

In Fig. \ref{entrain3d}, $\rho$ is the average density and $d$ is the
geometrical thickness of the outflow layer. In our case, the outflow is
geometrically thin ($d\ll l$).

$\Sigma$ is the integrated density, which is defined as
\begin{equation}
\Sigma=l_{\rm c} \rho= 2 \pi \omega \rho d\;, 
\end{equation} and $W$ is the
integrated momentum, which is defined as 
\begin{equation} 
W=l_{\rm c} \rho
v_{ x}= 2 \pi \omega \rho v_{ x} d\;, 
\end{equation} 
where $v_x$ is the
average outflow velocity along the outflow wall. 
Note that here $d$ is introduced for convenience. At a given position, we
calculated $\Sigma$ and $W$ first. $d$ can then be evaluated as 
$d=\Sigma/(2 \pi \omega \rho)$.

The equations of motion of the outflow layer describe the conservation of the
mass and the momentum,

\begin{eqnarray}  \label{equations:conservation}
\begin{split}
\frac{\partial}{\partial t}\Sigma(x,t) +\frac{\partial
}{\partial x}  \big ( v_{ x}(x,t) \Sigma(x,t) \big) &=& S_{\rm \Sigma}(x)\\ \frac{\partial}{\partial
t} W(x,t) +    \frac{\partial }{\partial x} \big(   v_{ x}(x,t) W(x,t) \big)      &=& S_{\rm W}(x,\Sigma)\;,
\end{split}
\end{eqnarray} where $W=v_{ x} \Sigma$. 

To solve it, we need the initial conditions $\Sigma(x,t=0)$ and $W(x,t=0)$.

We
still need to specify the source terms $S_{\rm \Sigma}$ and $S_{\rm W}$.
According to the geometry in Fig. \ref{balance}, $S_{\Sigma}$ can be
expressed as the sum of the mass flow from the wind and the mass flow from the
envelope,

\begin{equation}\label{eq2} S_{\Sigma}=S_{\Sigma \rm wind}+S_{\Sigma
\rm envelope}\;, 
\end{equation} 
where 
\begin{equation}\label{eq3} S_{\Sigma \rm
wind}= \rho_{\rm wind} v_{\rm wind}\sin \phi \times 2 \pi \omega\times \beta
\end{equation}
 and according to Equation \ref{entrain_turb},
\begin{equation}\label{eq4} S_{\Sigma \rm envelope}= \alpha
\rho_{\rm envelope} \sigma_{\rm turb} \times 2 \pi \omega\;. 
 \end{equation}
Here, $S_{\Sigma \rm wind} \ll S_{\Sigma\rm envelope}$. Most of the mass in the outflow comes from the envelope. 
 
$S_{\rm W}$
is the source term for the integrated momentum. $S_{\rm W}$
consists of the momentum injection by the wind, as well as the change of the momentum due
to gravity. $S_{\rm W}$ can then be expressed as,

\begin{eqnarray}\label{eq5} S_{\rm W} &=& S_{\rm W wind}+S_{\rm W gravity}\nonumber \\
&=& \rho_{\rm wind} v_{\rm wind}^2 \sin \phi \cos \phi \times 2 \pi \omega \times \beta +
\Sigma g\;, \end{eqnarray}

 where $g = G M_{*}/r^2 \cos \phi$ is the acceleration
due to gravity. Here, $M_{*}$ is the stellar mass (10 $M_{\odot}$), and $\cos \phi$ comes from
projecting the gravity onto the wall of the outflow cavity. $\beta$ is the wind efficiency.
It characterizes the efficiency with which the mixing layer absorbs
mass and momentum from the wind. We took 0.3 as its fiducial value.

\subsection{Local linear growth regime}
\label{sec:local_linear_regime}
Here we make an order-of-magnitude analysis of Equations 
\ref{equations:conservation}.
The first terms are the transient terms $\frac{\partial}{\partial t} \times P$.
The second terms are the advection terms, $\frac{\partial }{\partial x} (v_x
\times P) $, and the third terms are the source terms $S_{\rm P}$. Here, $P$ can
be either $\Sigma$ or $W$. If the outflow has initial mass and momentum
distribution $P_0= \Sigma_0$ \; or \;$W_0$, the mass/momentum accumulation
timescales can be written as
\begin{equation}\label{equation:accumulation}
t_{\rm accumulation}=P_{0}/S_{\rm P}\;.
\end{equation}
and the advection timescales can be written as
\begin{equation}
t_{\rm advection}=L/v_{\rm x}\;. 
\end{equation}
Here, L is the physical scale of the outflow, which is about 1 parsec. 

The mass/momentum accumulation timescale is dependent on the initial mass and
momentum $\Sigma_0$ and $W_0$. If the majority of the mass and momentum of the
outflow comes from the turbulence mixing process, $\Sigma_0$ and $W_0$ should be
insignificant, therefore $t_{\rm accumulation}$ is expected to be very short.
Because we lack the knowledge on the initial mass and momentum distribution of
the outflow, we assumed that the majority of the outflow mass and momentum comes
from the turbulence mixing process. This assumption is valid as long as the
outflow age is larger than the accumulation timescale $t_{\rm accumulation}$
(Equation \ref{equation:accumulation}) is long.
We studied the behaviour of Equations \ref{equations:conservation} by analysing
the importance of different terms. For simplicity, we neglected
the momentum added to the outflow due to self-gravity, therefore $W(x,\Sigma)$
in Equations \ref{equations:conservation} can be approximated as $W(x)$ .

The advection timescale $t_{\rm advection}$ is the time required for matter to
travel throughout the outflow. If the outflow has an age of $t_{\rm outflow}$,
the transient term is
\begin{equation}
{\rm TransientTerm}=\frac{\partial P}{\partial t}\sim \frac{P}{t_{\rm
outflow}}\;,
\end{equation}
and the advection term is
\begin{equation}
{\rm AdvectionTerm}=\frac{\partial v_x P}{\partial x}\sim \frac{v_x P}{L}\sim \frac{P}{t_{\rm advection}}\;.
\end{equation}
If we take the ratio between the advection term and the transient term, we have
\begin{equation}
\frac{\rm AdvectionTerm}{\rm TransientTerm}\sim \frac{v_x/L}{1/t_{\rm outflow}}\sim\frac{t_{\rm outflow}}{t_{\rm advection}}\;,
\end{equation}
where by definition $t_{\rm advection}=L/v$.
In our typical case, $t_{\rm outflow}=1 \times 10^4\rm yr$ and $t_{\rm
advection}\sim L/v_x\sim 1\; {\rm pc}/30\rm\; km\;s^{-1}\sim 1\times\; 10^5\rm yr$. $t_{\rm outflow}<t_{\rm advection}$, 
therefore advection is not important in the whole outflow.
In this case, the advection effect can smooth out the density variations at the small scale,
but can not alternate the structure of the whole outflow significantly.
This is also true for the outflow from the class 0 low-mass protostars
\citep{2007prpl.conf..245A}.

If the age of the outflow is small ($t_{\rm outflow}<<t_{\rm advection}$), as it
is most cases, we can neglect the advection terms in Equations
\ref{equations:conservation},  therefore we have
\begin{eqnarray}  
\frac{\partial}{\partial t}\Sigma(x,t) &=& S_{\rm \Sigma}(x)\\ 
\frac{\partial}{\partial t} W(x,t)  &=& S_{\rm W}(x)\;.
\end{eqnarray} 
The surface density of the outflow can then be calculated as
\begin{equation}\label{equations:transien}
\Sigma(x,t) =S_{\rm \Sigma}(x) \times t_{\rm outflow}\;,
\end{equation} 
and the velocity of the outflow can be calculated as 
\begin{equation}\label{equations:transient}
v_{\rm x}(x,t)=\frac{S_{W}}{S_{\rm \Sigma}}\;.
\end{equation}
In this case, the velocity profile of the outflow is fixed and the surface density of the outflow grows linearly with time. 

\subsection{Advection-term-dominated regime}
For the aged low-mass outflows (e.g outflows with $t\sim 10^7 \;\rm
yr$), we have $t_{\rm outflow} \sim t_{\rm advection}$, and the advection effect
can alter the mass distribution of the outflow significantly. Neglecting the
transient terms in Equations \ref{equations:conservation}, we have
\begin{eqnarray}  
\frac{\partial
}{\partial x}  \big ( v_{ x}(x,t) \Sigma(x,t) \big) &=& S_{\rm \Sigma}(x)\\ 
\frac{\partial }{\partial x} \big(   v_{ x}(x,t) W(x,t) \big)      &=& S_{\rm W}(x)\;,
\end{eqnarray} 
hence 
\begin{eqnarray}\label{equations:advection}
\begin{split}
v_x &=& \frac{\int S_{W}{\rm d} x}{\int S_{\rm \Sigma} {\rm d}x} \\
\Sigma &=& \frac{\Big(  \int S_{\rm \Sigma} {\rm d}x \Big)^2}{\int S_{\rm W} {\rm d}x}\;.
\end{split}
\end{eqnarray}
Here, both the velocity and the surface density exhibit stationary profiles.
By comparing Equation \ref{equations:transien} and \ref{equations:transient} with Equation \ref{equations:advection},
we find that when the outflow age is high, the advection term dominates
the transient term, which prevents the outflow surface density from
growing to infinity.

\subsection{Role of gravity}
\label{sec:gravity}
Here we consider the role of the gravity term in the structure of the outflow.
According to Equation \ref{eq5}, the contribution of gravity is quantified by the ratio
\begin{equation}
f_{\rm gravity}=\frac{\Sigma g}{\rho_{\rm wind} v_{\rm wind}^2\times 2\pi\omega\times \beta}\sim\frac{\rho d 2 \pi \omega g \alpha}{\rho_{\rm wind}\times v_{\rm wind}^2 \times 2 \pi \omega \beta}  \;,
\end{equation}
where $d=\sigma t \alpha$ is used. Here, $d$ is the thickness of the outflow,
$\alpha$ is the entrainment efficiency, $\sigma$ is the velocity dispersion of the envelope and $t$ is the outflow age. 
From this we have 
\begin{equation}
f_{\rm gravity}= \frac{g t}{\sigma}\times \frac{\alpha}{\beta}\;.
\end{equation}
Here, $\sigma$ is the velocity dispersion of the envelope, $t$ is the age of the
outflow, and $g\sim G M_{*}/r^2$ is the local gravity.
Since $g\sim r^{-2}$, gravity will be important at regions close to the protostar, and it will be important when the mass of the outflow has grown significantly. 

Considering the inner region of the outflow, we have $M_{*}=10 M_{\odot}$,
$r=0.1 \rm\; pc$ and $t=10^4\;\rm yr$. Therefore, $f_{\rm gravity}\sim 10^{-2}$
and the gravity from the protostar is negligible. However, at regions closer to
the protostar, gravity must play an important role in slowing down the outflow.
\subsection{Numerical results}

To illustrate how the mass  and the momentum of the outflow grows with time, we
assume a static envelope and a wind with a constant mass-loss rate and followed
the evolution of the entrainment layer by solving Equations
\ref{equations:conservation} -- \ref{eq5} numerically. The cavity shapes are
calculated in \S \ref{cavity_shape}.  At $x=0$, we imposed
\begin{eqnarray}
\frac{\partial \Sigma}{\partial x}&=&0\\ \frac{\partial W}{\partial x}     &=&0 \;,
\end{eqnarray}
and set the initial condition so that
 \begin{eqnarray} 
 \Sigma(x)  &=&   S_{\Sigma}
\Delta t_0\\ v_x(x)    &=&   1\times 10^{-4}\rm\; cm\; s^{-1}\;,
 \end{eqnarray} 
 
 where $\Delta t_0= 100\; yr$.
The equations are solved using FiPy (J. E. Guyer, D. Wheeler \& J. A.
Warren)
\footnote{Avaliable at http://www.ctcms.nist.gov/fipy/}. 

We are mainly interested in the transient-term-dominated regime, since because
this regime is relevant to all high-mass protostellar outflows and the
majority of the young low-mass protostellar outflows.

\begin{figure*}
\includegraphics[width=0.9 \textwidth]{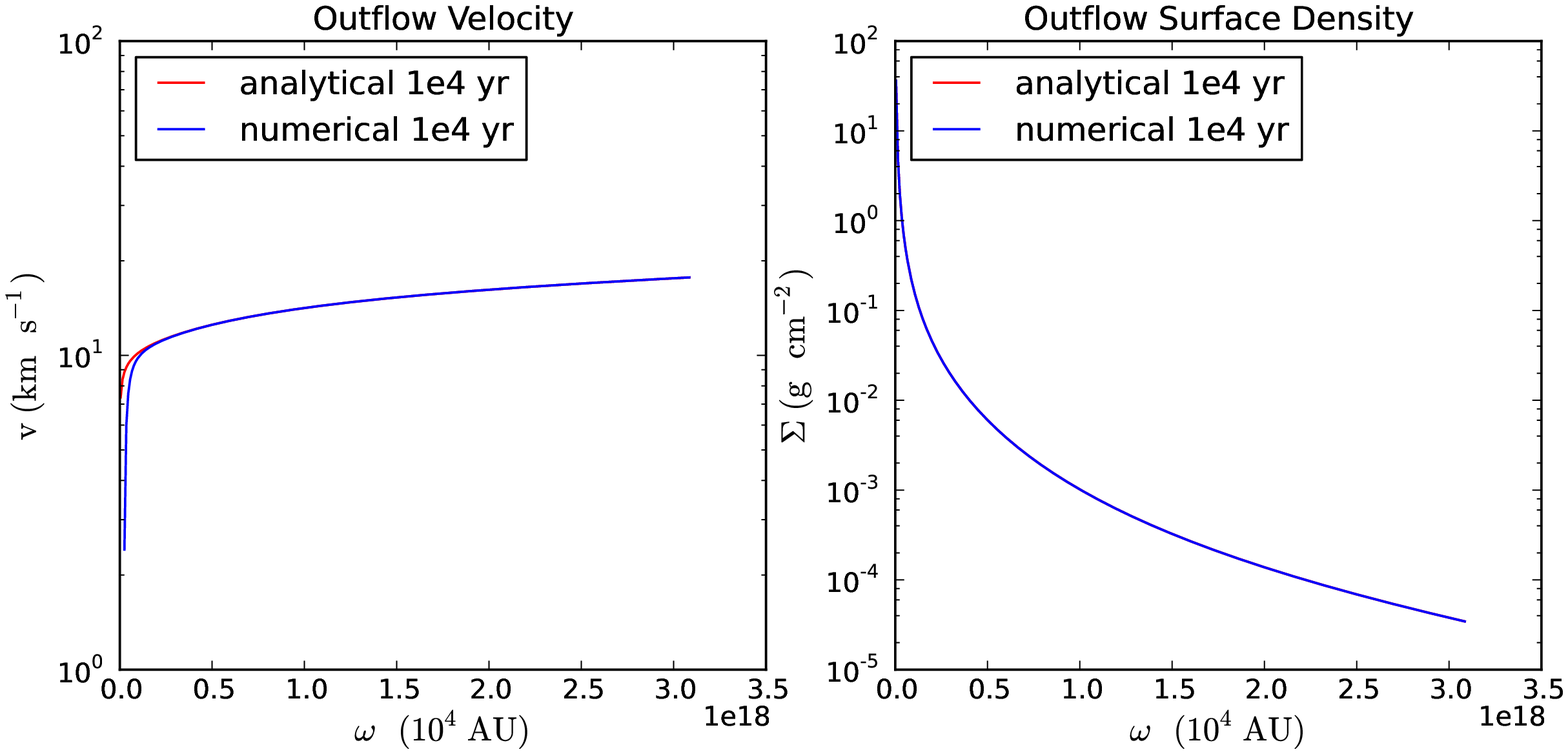}\\
\includegraphics[width=0.9 \textwidth]{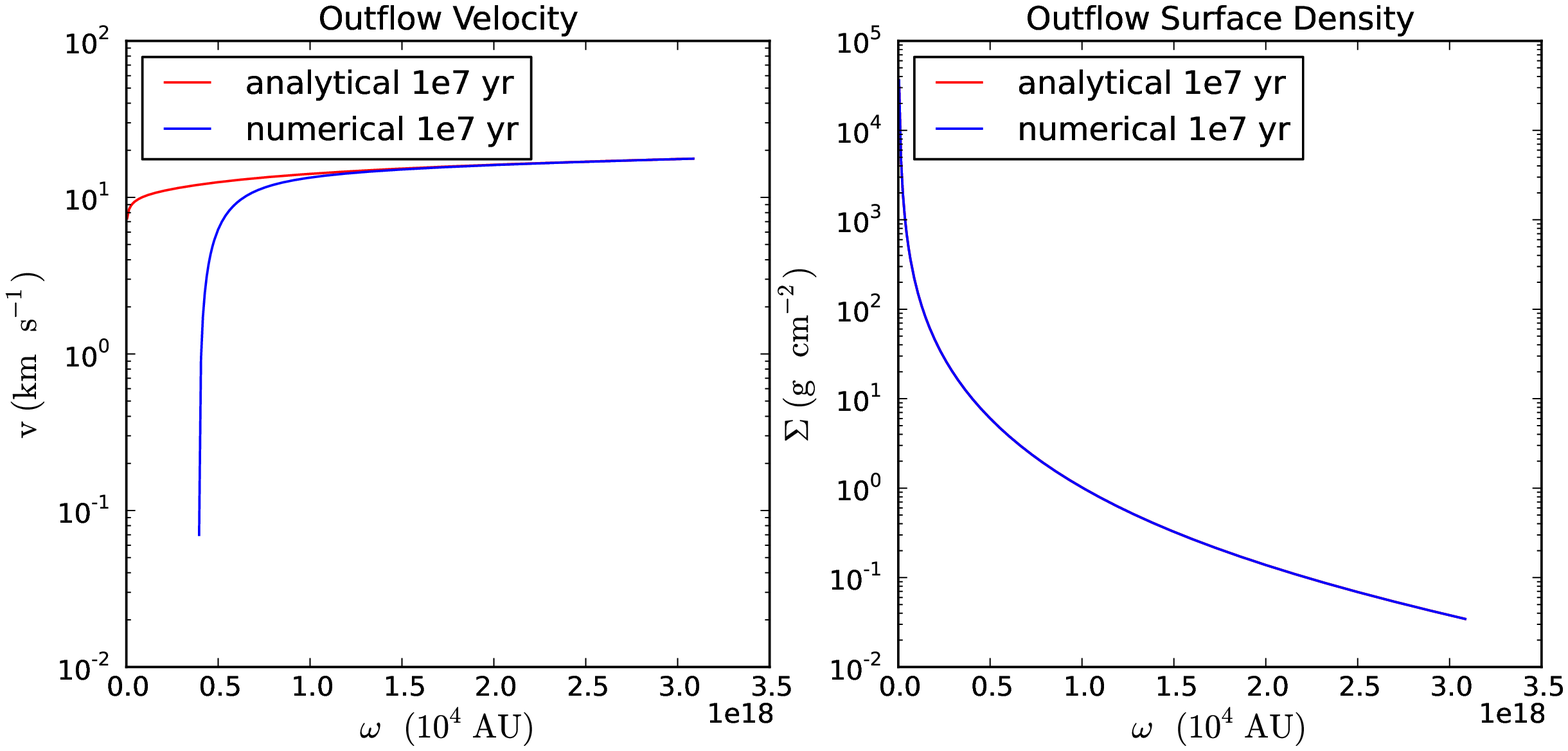}
\caption{Velocity and surface density distribution of the outflow calculated
by solving Equations \ref{equations:conservation}.
The blue lines are the numerical results and the red lines are results calculated by assuming the local
conservation of energy and momentum (\S \ref{sec:local_linear_regime}).
The upper panels are the results for a $10^4 \;\rm yr$ outflow.
The lower panels are the results for an unrealistically old ($10^7\;\rm yr$)
outflow.
At smaller radii, the deviation of the numerical solutions from the analytical
solutions are caused by the gravity term in Eq. \ref{eq5}.
\label{figure:profiles}}
\end{figure*}

In Fig. \ref{figure:profiles}, we plot the results from our calculations for a
realistic $10^4\;\rm yr$ outflow and unrealistically old $10^7\; \rm yr$
outflow. In our $10^4\;\rm yr$ outflow, the numerical solution agrees quite well
with the analytical solutions (\S \ref{sec:local_linear_regime}). At small
radius, the numerical solutions show lower velocity. This is because the
matter in regions close to the protostar is slowed down by the gravity from the
protostar.

We also plot the results of an outflow with an unrealistically long
lifetime of $10^7\;\rm yr$. Because here gravity plays a much more
significant role in slowing down the outflow (\S \ref{sec:gravity}), the
velocity of the outflow at the central part of the outflow becomes much lower.

In most cases, the effect of gravity is negligible in large portions of the
outflow,  the analytical solutions can be taken as good approximations to the
structure of the outflow. However, it is difficult for the outflow to enter the
advection-dominated regime, since here gravity tends to dominate the velocity
structure of the outflow and to slow down the outward motion before advection is
able to change the structure of the whole outflow significantly.

The final state we obtained from solving the equations of mass and momentum
conservation in the {\it local linear growth regime} is representative of the
real structure of the outflow, as long as the mass and momentum added to the
outflow is much larger than the initial mass and momentum of the outflow. If
the strength of the wind or the structure of the envelope changed during the
evolution of the system, the final state of the outflow will change accordingly.
Therefore, the system will evolve towards a new state, and the time for this
evolution to take place is again the accumulation timescale (Equation
\ref{equation:accumulation}). The older the outflow becomes, the more difficult
it is for the outflow to relax to the new state.

In our present work, we assumed that the wind and envelope remain stationary.
Our solution is valid when the structure of the wind and the structure of the
envelope change slowly. Since detailed observations of the way in which the wind
and envelope evolve are currently unavailable, we assume this simplicity. If the
wind and the envelope both evolve significantly, it is still possible to
calculate the structure of the outflow by making some modifications to the
framework presented.

\subsection{Structure of the entrainment layer} 
\label{sec:entrain}
\begin{figure}
\includegraphics[width=0.42\textwidth]{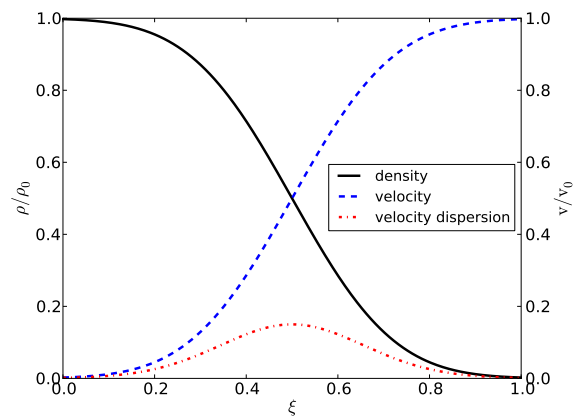}
  \caption{Density and velocity structure of the entrainment layer.
  The (black) solid line denotes the density structure (equation \ref{eq:density}),
  the (blue) dashed line denotes the velocity structure (Equation
  \ref{eq:velocity}), and the (red) dotted line denotes the structure of the fluctuating velocity (equation \ref{eq:random_v}).
  $\xi$ is defined in Equation \ref{eq:eta}. \label{fig:entrainment_inside}}
\end{figure}

To fully specify the structure of the outflow, we need to know the structure of
the entrainment layer. The gas motion inside the entrainment layer is dominated
by turbulence, and its velocity can be expressed as the sum of a mean component
and a fluctuating component \citep{1895RSPTA.186..123R}.
There are many experiments and numerical simulations that study the structure of
such a mixing layer
\citep[e.g.][]{1976JFM....74..209C,1990AIAAJ..28.2034B,1994PhFl....6..903R},
from which we obtained the structures of the mean velocity and the fluctuating
(random) velocity, and parametrized them.

 We found that the structure of the entrainment layer is universal and can be
 represented in a simple parametrized way. We fitted the mean density, the mean
 velocity and the turbulent velocity structures of \citet{1994PhFl....6..903R},
 and applied the fitting formula to our case.
 
 The fitted density and mean velocity distributions take the forms of

\begin{equation}\label{eq:density}
 \rho(\xi)= \frac{
-{\rm erf} (4 \xi -2) +1}{2}  \rho_{\rm envelope}\;,
 \end{equation} 
and
\begin{equation}\label{eq:velocity}
v(\xi)= \frac{ {\rm erf} (4 \xi -2) +1}{2}  v_{\rm
norm}\;,
\end{equation} 
where $\rm erf$ is the error function, and $\xi$ is
defined as 
 \begin{equation}\label{eq:eta}
\xi=\frac{x}{2 d}\;.  
 \end{equation}
 $d$ is defined as $d=\Sigma\ /\rho_{\rm envelope}$, and $v_{\rm
norm}=5 \times v_{ x}$ due to the normalization requirement.

The distribution of the fluctuating velocity of the entrainment layer takes the
form
\begin{eqnarray}\label{eq:random_v}
 v_{\rm turb} &=& v_{\rm turb\_norm}
({\rm erf} (4 \xi -2) +1)( -{\rm erf} (4 \xi -2) +1)\nonumber \\ &=& v_{\rm
turb\_norm}  (1 -{\rm erf}^2(4 \xi -2)) \;,
 \end{eqnarray} 
where $v_{\rm
turb\_norm}=0.15 \times v_{\rm norm}$.  
 Fig. \ref{fig:entrainment_inside} shows the structure of density,
 mean velocity, and fluctuating velocity inside the entrainment layer.

\begin{figure}
\includegraphics[width=0.48\textwidth]{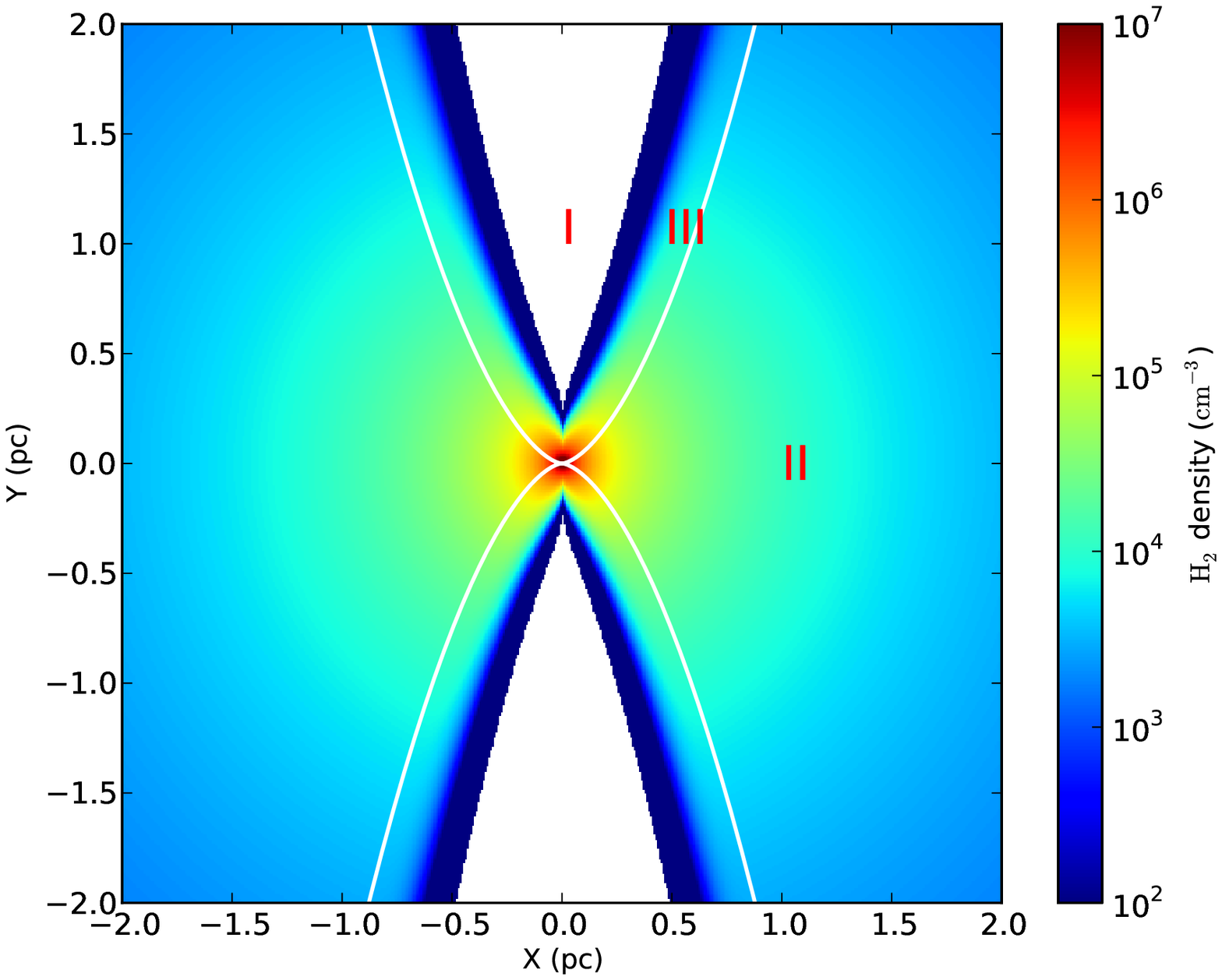}
 \caption{Density structure of the outflow. Region \Rmnum{1} represents the
 inner cavity blown by the wind, region \Rmnum{2} represents the envelope, and
 region \Rmnum{3} represents the outflow layer. \label{fig:density} The wind
 mass-loss rate is $1.5\times 10^{-3}\;\rm M_{\odot}\; yr^{-1}$.
 The thickness of the entrainment layer has been scaled up by a factor of $30$
 for clarity. }
\end{figure}

Fig. \ref{fig:density} shows the density distribution of our model, from which
the entrainment layer can be identified as the region between the outflow cavity
and the envelope.

\section{Observational tests}\label{sec:obs}
\begin{figure*}

\includegraphics[width=0.95\textwidth]{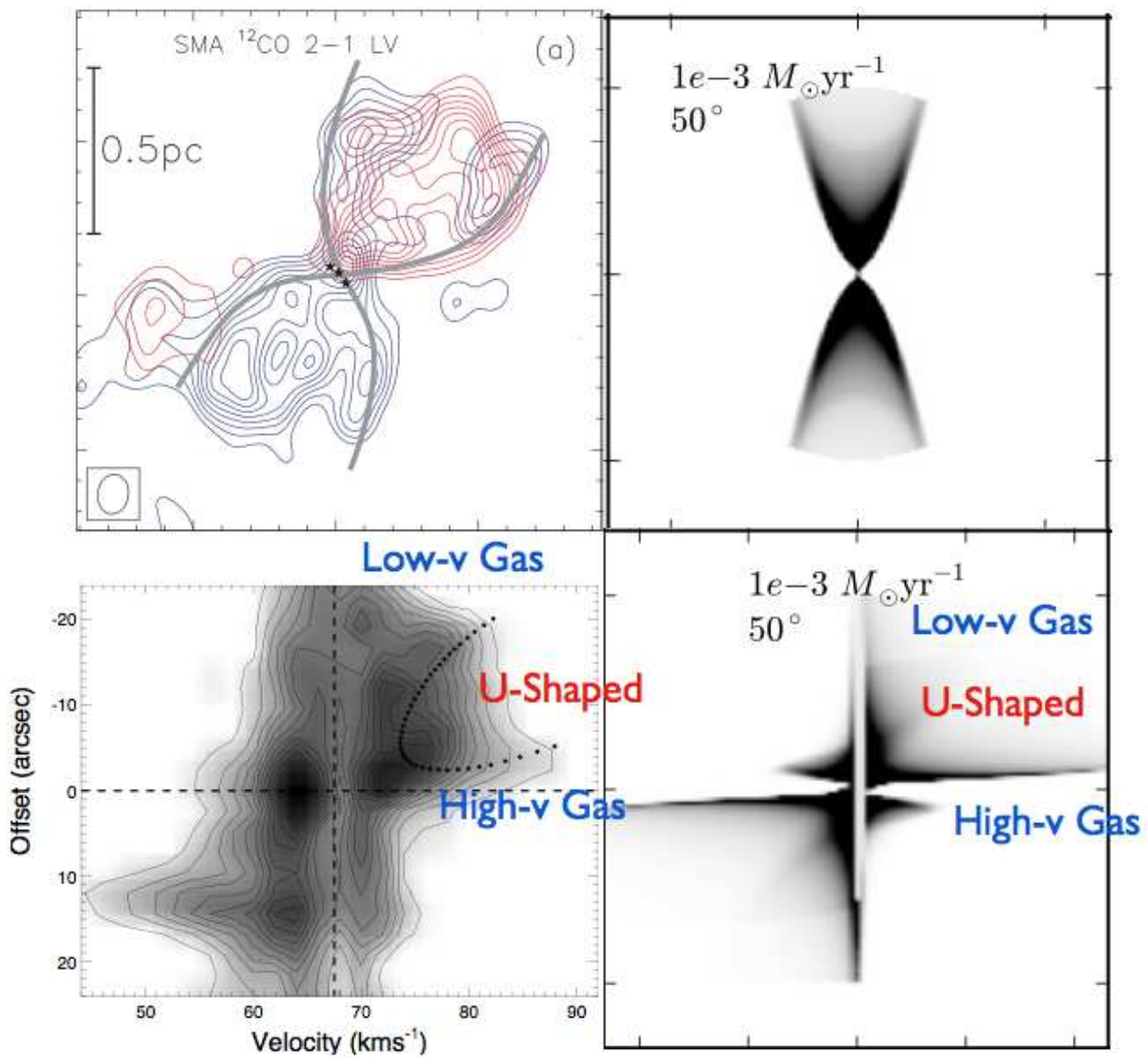}

  \caption{ 
  Comparison between the observations by \citet{2009ApJ...696...66Q} and the
  predictions from our wind-driven turbulent entrainment model.
The {\bf upper left panel} shows the morphology of the outflow observed by
\citet{2009ApJ...696...66Q}, the {\bf lower left panel} shows the
position-velocity structure of the observed outflow.
The {\bf upper right panel} shows the morphology of the outflow from our
modelling, the {\bf lower middle panel} shows the position-velocity
structure of our outflow model cut along its major axis.
The outflow model has a mass-loss rate of $1.0\times 10^{-3} M_{\odot}\;\rm
yr^{-1}$ and an inclination angle of $50^{\circ}$.  For both the observations
(lower left panel) and our modelling (lower right panel), the U-shaped
region is the region in the position-velocity diagram where the structure of the
outflow exhibits a U shape, the Low-v gas region is the region where
the gas has a relative small velocity at regions far from the protostar, and the
High-v gas region is the region where the gas has a relatively high velocity
in the close vicinity to the protostar.
 \label{fig:image_spec}}
\end{figure*}

\label{sec:obs}

\begin{figure*}
\includegraphics[width=0.95\textwidth]{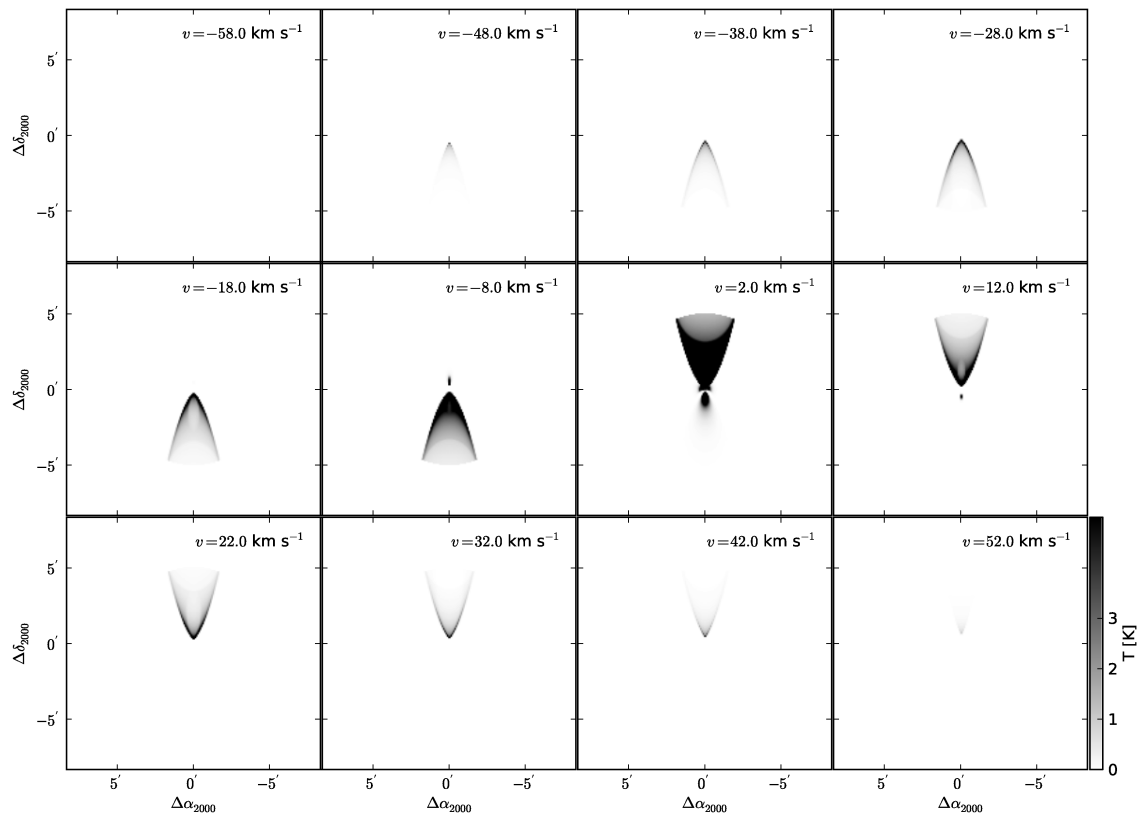}
  \caption{CO(3-2) channel map of an outflow that has a mass-loss rate of
  $1.0\times 10^{-3} M_{\odot}\;\rm yr^{-1}$ and an inclination angle of $50^{\circ}$.\label{fig:channel_map}}
\end{figure*}

\begin{figure*}
   \centering
\includegraphics[width=0.95\textwidth]{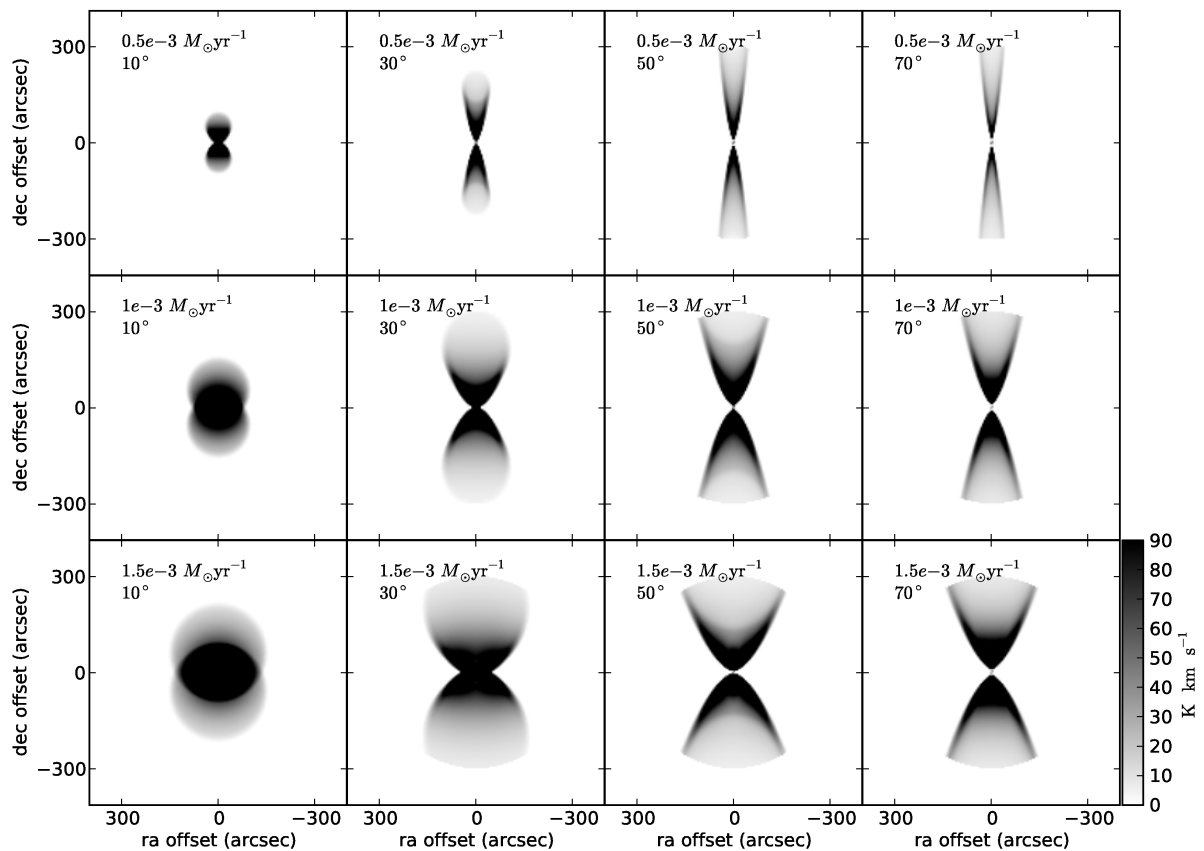}
  \caption{  Velocity-integrated CO(3-2) images of outflows from our model with
  different wind mass-loss rates and inclination angles.  \label{fig:grid_image}}
\end{figure*}

\begin{figure*}
\includegraphics[width=0.95\textwidth]{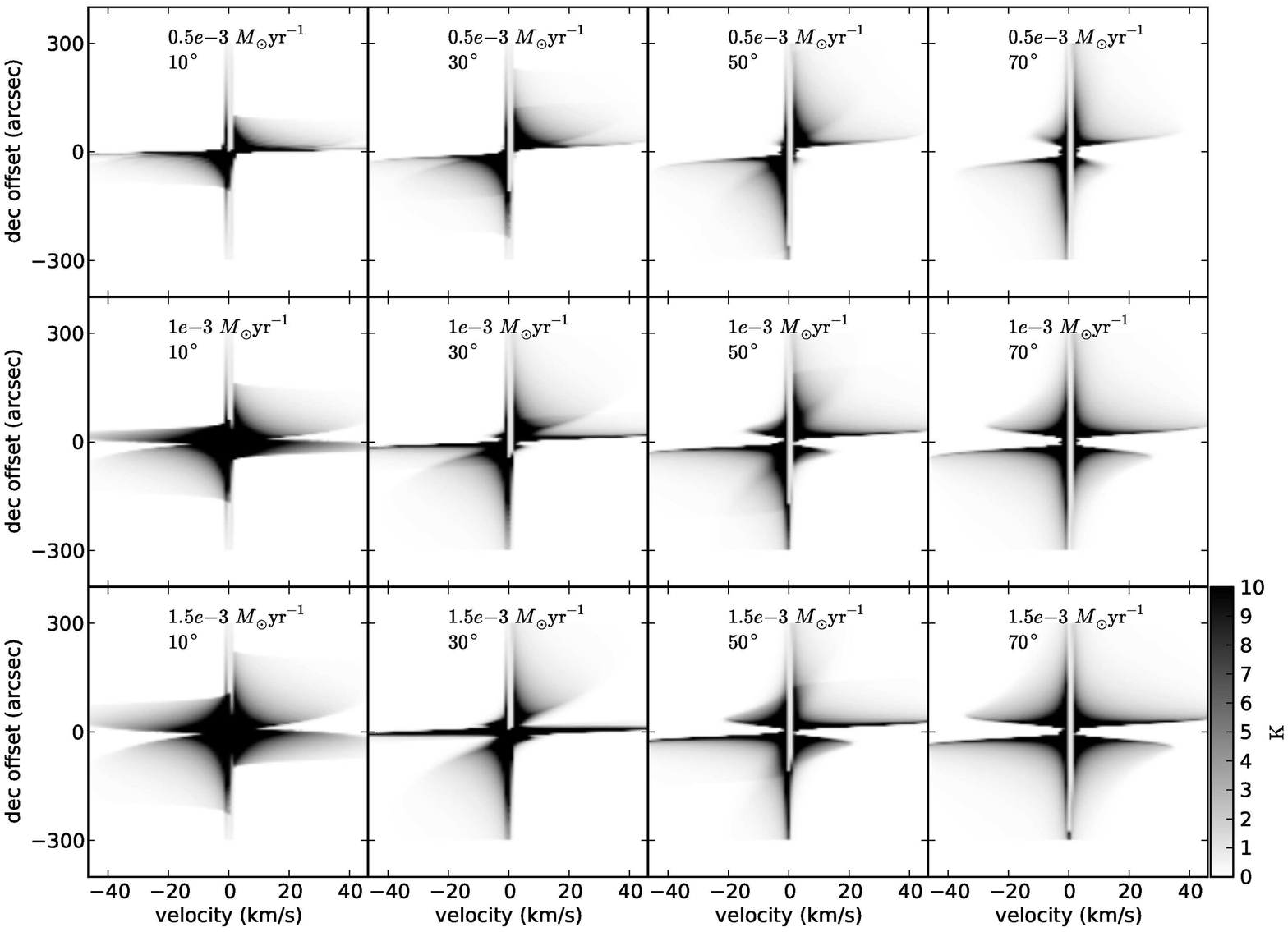}
  \caption{CO(3-2) position-velocity diagram cut along the major axis of our outflow models with diferent wind mass loss rates and inclination angles. \label{fig:grid_pv}}
\end{figure*}

Observationally, molecular outflows are frequently mapped in the emission from
rotational transitions of molecules such as CO and HCO$^+$.
These mapping observations can measure the intensity of the line emission from
the outflow in the form of three-dimensional
xyv ($x$-position-$y$-position-velocity) data cubes.
CO is the most abundant and widespread molecule in the interstellar medium apart
from the difficult-to-observe $\rm H_2$, and has been used to trace the bulk of
molecular gas. CO observations \citep[][and references
therein]{1985ARA&A..23..267L,2007prpl.conf..245A} have revealed a diversity of
molecular outflows, some of which show remarkably regular morphologies
\citep[e.g.][]{2009ApJ...696...66Q}.

Here we concentrated on modelling the emission from our outflows and studied the
connection between the physical structure of the outflow models and the
structure of our outflow models observed in a 3D data cube
\citep[e.g.][]{1986ApJ...307..313C,1990ApJ...348..530C,1992A&A...261..274C,1994ApJ...422..616S}.
To model the emission from the outflow, we need to know its dynamics and its
abundance/excitation conditions.
The dynamical structure of the outflow has been obtained in \S
\ref{sec:model}. However, the abundance and excitation are still uncertain:
 in regions close to the central protostar, the CO molecule can be
 photodissociated and cannot trace the gas anymore. The temperature of the
 outflow layer is determined by the balance between various heating and cooling
 processes, which are also uncertain.
While these effects will influence the spacial and velocity distribution of the
observed CO emission, the morphology of the outflow as well as its overall
structure in the position-position and position-velocity spaces are relatively
unchanged.
Therefore we assume that the CO abundance is
$10^{-4}$ relatively $\rm H_{2}$ and that the outflow entrainment layer has a
constant kinematic temperature of $\sim 100\rm K$.
In the following discussions, we focus on the overall morphology of the
outflow in the position-position and position-velocity spaces.

For the velocity dispersion, velocity gradient and kinematic temperature from
our model, we used  a python version of RADEX \citep{2007A&A...468..627V} to
calculate the population distribution of the CO molecule.
We then used a ray-tracing code to calculate the line emission from our
outflows.
 Our ray-tracing calculations were made with the help of LIME
\citep{2010A&A...523A..25B}.

Fig. \ref{fig:image_spec} shows a comparison between our model and the
observations by \citet{2009ApJ...696...66Q}.
We show the integrated image of the outflow and the position-velocity diagram of
the outflow cut along the major axis. The integrated images of the outflow show
a regular conical shape in both the observations and our modellings, and the
position-velocity diagram of the outflows from observations and modellings show
similar shapes.
In regions close to the protostar, the observations show a velocity dispersion
of about $20 \rm \;km \;s^{-1}$. This velocity dispersion comes from a
continuous distribution of emission and is consistent with observational
studies \citep[e.g.][]{2011ApJ...728....6Q,2011ApJ...729..124C}.

We propose that this feature is one strong evidence for the existence
of the entrainment layer. By looking at one position in the image, we integrate
through the whole entrainment layer. In one line of sight, this produces a
continuous distribution of fluids that move at different speeds (Figure
\ref{fig:entrainment_inside}). Seen from the spectral line profile, we can
always identify a broad component, since the line emission traces the
distribution of the mass, and the mass distribution inside the entrainment layer
is continuous.

On the other hand, the observation shown in Fig. \ref{fig:image_spec} is
difficult to understand in the context of the wind-driven shell model.
In that model, the velocity of the outflow is proportional to the distance from
the protostar and should vanish at close vicinity of the protostar, which is
not observed (Fig. \ref{fig:image_spec}). {Moreover, the outflow speed are
high at regions far away from the protostar because only a high expansion
velocity can make the gas move far.} But in the observations we
can still see low-velocity gas in regions far from the protostar.

Figure
\ref{fig:channel_map} 
shows a channel map of an outflow with a mass-loss rate of $1.0\times 10^{-3}
M_{\odot}\;\rm yr^{-1}$ and an inclination angle of $50^{\circ}$. The inner part
of the outflow is visible in most channels, implying a broad velocity spread at
this location. This velocity spread is a direct consequence of the turbulent
entrainment process.

Figures \ref{fig:grid_image} and \ref{fig:grid_pv} show the calculated images
and position-velocity diagrams of outflows from protostars with different mass-loss
rates viewed at different inclination angles. The mass-loss rate takes
values of 0.5, 1, 1.5, and $\times 10^{-3}\;\rm M_{\odot} \;yr^{-1}$, the
inclination angles values of $10^{\circ}$, $30^{\circ}$, and $50^{\circ}$. The
calculated outflows exhibit a variety of morphologies. They also share some
common characteristics: The outflow gas can
reach a high velocity in regions close to a protostar; in regions far away from
the protostar, there is still gas with low velocities. These features are natural outcomes of the entrainment
process  (\S \ref{sec:entrain}) and can be observed under different
circumstances.

\section{Outflow entrainment as a universal process}\label{sec:discu}
Outflows  are ubiquitously observed in a variety of situations, which include
the formation of low-mass stars and high-mass stars, and the
situation in which a wind blown by an AGN interacts
and entrains the galactic-scale ambient gas
\citep{2011ApJ...735...88A,2012ApJ...752...38T}. Here we considered the
possibility that these outflows have a common origin, in that all these outflows
are formed through the interaction between the wind and the ambient gas in the
form of turbulent entrainment.

In \S \ref{sec:collimation}, using the turbulent core model of massive star
formation \citep{2002Natur.416...59M,2003ApJ...585..850M}, we show that outflows
from low and high-mass protostars can
consistently be interpreted as resulting from an universal entrainment process. In \S
\ref{sec:massvelo}, we derive universal scaling relations that can be used to
estimate the mass and velocity of the outflows, and suggest that the
AGN-driven outflows can be consistently explained in our model. In \S
\ref{sec:drawf}, using the insights obtained from the scaling analysis, we
predict the existence of dwarf outflows in cluster-forming regions, and in 
\S \ref{sec:selfsim} we discuss the self-similarity of our model.

\subsection{Universal picture of protostellar outflows}\label{sec:collimation}

As suggested by \citet{2002Natur.416...59M,2003ApJ...585..850M}, the formation
of high-mass stars can be viewed as a scaled-up version of the standard low-mass
star formation theory from gas cores, where most of the pressure support is due
to a combination of turbulence and magnetic fields instead of thermal pressure.
In this picture, a high level of
turbulence prevails in the high-mass star forming regions, causing significant
turbulent ram-pressure of the ambient medium. The timescale for the formation of high-mass stars is short, and the
accretion rate onto the high-mass protostar is much higher than in the low-mass
case.

Our model of a turbulent entrainment outflow is self-similar in nature. By
combining it with the self-similar star formation model, we can obtain a
universal description of outflows from low and high-mass
protostars.

Considering the fiducial case in \citet{2003ApJ...585..850M}, the accretion rate
onto the protostar can be estimated as
\begin{equation}\label{eq:md}
\dot m_{*}\sim 4.6\times10^{-4}\times \Big( \frac{m_{*f}}{30\; M_{\odot}}\Big)^{3/4}\times \Sigma_{\rm cl} \times \Big(  \frac{m_{*}}{m_{*f}}  \Big)^{0.5} M_{\odot}\; \rm yr^{-1}\;,
\end{equation}
where $m_{*}$ is the current mass of the protostar, $m_{*f}$ is the final mass of the protostar when the accretion has been finished, and $\Sigma_{\rm cl}$ is the surface density of the clump.
High-mass star-forming regions are characterized by a high value of $\Sigma_{\rm
cl}$ \citep{2003ApJ...585..850M}.
The pressure of the wind from the protostar can be estimated as
\begin{eqnarray}\label{eq:pwindk}
\frac{p_{\rm wind}}{k_{\rm B}}&=&\frac{\dot m_{\rm wind}\times v_{\rm wind}}{\pi r^2} \nonumber \\
&=& 0.77 \times 10^{8} \times \Big( \frac{m_{*f}}{30\; M_{\odot}}\Big)^{3/4}\times \Sigma_{\rm cl}^{3/4}\times \nonumber\\
&\;& \Big(\frac{r}{\rm pc}  \Big)^{-2}\times  \Big(  \frac{m_{*}}{m_{*f}}  \Big)^{0.5} \;\rm K\; cm^{-3} \;,
\end{eqnarray}
where $k_B$ is the Boltzmann constant, and we assumed $\dot m_{\rm
wind}=1/3\times \dot m_{*}$ \citep[][where $\dot m_{*}$ is the total accretion rate onto the protostar]{1994ApJ...429..808N} and $v_{\rm wind}=600\;\rm km\;s^{-1}$.  The average pressure of the clump is expressed as \citep{2003ApJ...585..850M}
\begin{equation}\label{eq:penvelopek}
 \frac{\overline p_{\rm cl}}{k_{\rm B}}=\frac{0.88 \;G}{k_{\rm B}} \; \Sigma_{\rm cl}^2=4.2\times 10^8\; \Sigma_{\rm cl}^2 \;\rm K\; cm^{-3}\;.
\end{equation}

For typical parameters, the average pressure of the envelope $\overline p_{\rm
cl}$ is similar to the average pressure of the wind at $1\;\rm pc$, which is
approximately the physical size of the outflow. This justifies our suggestion
that the ram-pressure of the wind and the ram-pressure of the ambient medium are
similar.

The collimation of the outflow is determined by the ratio between the turbulent
ram-pressure of the envelope and the ram-pressure of the wind. Stronger winds
lead to less-collimated outflows while more turbulent envelopes lead to highly
collimated outflows. To summarize these effects, we define the {\it dimensionless} collimation parameter $\eta$ of the outflow:
\begin{eqnarray}\label{eq:collimation}
\eta &=& \frac{\overline p_{\rm cl}}{p_{\rm wind}}\ \nonumber \\
 &=&  5.5 \times  \Big(  \frac{r}{\rm pc}  \Big)^{2} \Big( \frac{m_{*}}{m_{*f}}\Big)^{-1/2} \Sigma_{\rm cl}^{1/2} \Big(\frac{m_{*f}}{30 M_{\odot}} \Big)^{-3/4}\;,
\end{eqnarray}
with larger $\eta$ implying more collimated outflows. 

Equation (\ref{eq:collimation}) has several implications.

The ratio $\eta$ is proportional to $r^{-2}$, which means that the pressure of
the envelope gradually dominates the pressure of the outflow tends to
collimate the outflow with increasing distance from the protostar. This may
explain why many outflows exhibit a re-collimated shape, that is, the
opening angle of the outflow becomes smaller as we moves away from the protostar \citep[e.g.
L1157, ][]{1997ApJ...487L..93B}.

Also, $\eta \sim \Sigma_{\rm cl}^{1/2}$, which means that higher external
pressure (stronger turbulence) leads to more collimated outflows. This agrees
with the results in \S \ref{sec:shape}, and indicates that turbulence can {\it collimate} the outflow.

Third, if the pressure of the clump $\Sigma_{\rm cl}$ is roughly constant, the
more massive the star is, the less collimated is the outflow. This is because
$\eta$ depends on the final mass of the protostar, $\eta \sim m_{*f}^{-3/4}$.
The more massive the star, the stronger the wind it has. This stronger wind
will push the envelope and leads to a less collimated outflow. If several stars
form in a clustered way \citep{2011ApJ...728....6Q}, protostars of different
mass will share one common environment. We then expect that more massive
protostars produce less-collimated outflows.

\subsection{Outflow mass and velocity}\label{sec:massvelo}
According to our model, the mass of the outflow is determined by the efficiency
with which the outflow entrains the envelope gas.
Given the opening angle, the mass of the outflow can be estimated as
\begin{equation}\label{eq:totalmass}
M_{\rm outflow} \sim \alpha\; \rho_{\rm average} \; \sigma_{\rm average} \times t \times L^2\;.
\end{equation}
where $L$ is the physical scale of the outflow and $\rho_{\rm average}$ and $\sigma_{\rm average}$ are the average density and velocity dispersion, respectively. 
The average pressure of the envelope is 
\begin{equation}\label{eq:penvelope}
p_{\rm envelope}\sim\rho_{\rm average}\times \sigma_{\rm average}^2\;.
\end{equation}

The momentum of the outflow is determined by the amount of momentum injected by the wind and can be estimated as
\begin{equation}\label{eq:totalmomentum}
P_{\rm outflow}\sim \dot M_{\rm wind}\times t \times v_{\rm wind} \times \beta \;,
\end{equation}
and the ram-pressure of such a wind can be estimated as
\begin{equation}\label{eq:pwind}
p_{\rm wind} \sim \frac{ \dot M_{\rm wind}  \times v_{\rm wind}  }{L^2}\;.
\end{equation}

If the cavity blown by the outflow is stable, the ram-pressure of the wind
$p_{\rm wind}$ is expected to be similar to the turbulent ram-pressure of
the envelope, $p_{\rm envelope}$, and we have (from Equation \ref{eq:penvelope} and
Equation \ref{eq:pwind})
\begin{equation}\label{eq:forcebal}
\dot M_{\rm wind} \times v \sim \rho_{\rm wind} \sigma_{\rm wind} L^2 \; .
\end{equation}

The outflow velocity is then (equations \ref{eq:totalmass}, \ref{eq:penvelope}, and \ref{eq:forcebal})
\begin{eqnarray}\label{eq:outflowv}
v_{\rm outflow} &\sim& \frac{P_{\rm outflow}}{M_{\rm outflow}}\sim \frac{\dot M_{\rm wind} v_{\rm wind}}{\rho_{0} \sigma_{0} }\times \frac{\beta}{\alpha}\nonumber  \\
 &\sim& \sigma_{\rm average} \frac{\beta}{\alpha} \sim 3\times \sigma_{\rm average}\;.
\end{eqnarray}\\
In the last step we inserted the numerical values of $\alpha$ and $\beta$.
The velocity of the outflow is several times higher than the velocity dispersion
of the envelope.

This fact can be understood as follows: If the wind and the envelope can
establish hydrostatic equilibrium, then increasing turbulent
velocity $\sigma$, the pressure of the wind has to scale according to $\sigma^2$
 to balance wind pressure. Therefore, the momentum injection
of the wind scales as $\sigma^2$. On the other hand, the mass supply of the
outflow from the envelope in the form of the entrainment process only scales as
$\sigma^1$. The outflow velocity, which is estimated as $P/M$, scales as
$\sigma^1$.

These results outline one important property of the turbulent entrainment
outflow, namely that the velocity of the outflow is higher than while still
similar to the velocity dispersion of the ambient gas. This is independent of
other parameters such as the strength of the wind, and is based only on the assumption that the wind and the envelope can establish hydrostatic equilibrium.
{ In recent {\it Herschel}-HIFI
 observations of high-J CO emission in low and high-mass star-forming
 regions \citep{2013arXiv1301.4658S}, it was found that the velocity of the
 outflows (traced by the FWHM of the broad $^{12}$CO(10-9) line emission) is
 higher than while still similar to the velocity of the envelope (traced by
 the FWHM of the C$^{18}$O(9-8) line).
Also interesting is the fact that broader envelope line width is usually
associated with broader outflow line width. This implies that the outflows
from different regions share some physical similarities, and one possibility is
that all these outflows come from the same entrainment process.}

In a recently discovered AGN-driven outflow \citep[][]{2011ApJ...735...88A}, the
line width of the ambient gas (single-peaked central velocity component) is
$\sigma_{\rm envelope}\sim 100 \; \rm km \; s^{-1}$, while the width of the
broad line wing caused by the outflow activity is $v_{\rm outflow}\sim 300 \;\rm km \;
s^{-1}$. Although velocity of the ambient gas and the
velocity of the outflow are much higher than those of protostellar outflows,
their ratio $v_{\rm outflow}/\sigma_{\rm envelope}$ is about 3. This is
similar to the ratio predicted in our model (equation \ref{eq:outflowv}), which
suggests that these outflows may be the outcome of the turbulent entrainment
process working between the wind emitted by the central AGN and the
galactic-scale ambient gas.

\subsection{Dwarf Outflows}\label{sec:drawf}

As we have shown in \S \ref{sec:collimation}, the emergence of an outflow
depends on the pressure of the envelope and the pressure of the
wind reaching a balance. This is not always true. If the wind from the
protostars is not strong enough to blow away the envelope, we expect to see that
the outflows have small spatial extent and are confined to small regions.

Such ``Dwarf'' outflows may exist under several conditions. If several
protostars form in a clustered way \citep[e.g.][]{2011ApJ...728....6Q}, then the
outflows from the low-mass protostars of different mass may be confined by the
turbulence and appear as ``dwarf outflow''. Alternatively, such ``Dwarf''
outflows may exist at regions where the wind from the protostar is extremely
weak \citep[e.g. VeLLOs][]{2009ApJ...693.1290L,2010ApJ...721..995D}. Studies of
such ``Dwarf'' outflows will help to gain insights into the interaction between
the wind and the envelope in extreme cases.
\subsection{Self-similarity of the model}\label{sec:selfsim} 
We explored the self-similar properties of our model. Each model is
characterized by several parameters: a wind mass-loss rate $\dot M_{0}$, a
turbulence velocity $\sigma_{ 0}$, a wind velocity $v^{\rm wind}_{0}$, a density
$\rho_{0}$, an outflow age $t_{0}$, and the entrainment parameters $\alpha_{0}$
and $\beta_{0}$. { Our model is self-similar if when gravity is neglected.}
Here we derive the condition at which the structure of another outflow
characterized by a different set of parameters  ( $\dot M_{1}$, $\sigma_{1}$,
$v^{\rm wind}_{1}$, $\rho_{1}$, $t_{1}$, $\alpha_{1}$, and $\beta_{1}$ ) can be
obtained from scaling of a different outflow.

The opening angle of the outflow is determined by the relative strength of the
ram-pressure of the wind and the turbulent ram-pressure of the envelope. For it
to be unchanged, we require
\begin{equation}\label{eq:scaling1}
\frac{\rho_{1}\sigma_{1}^2}{\rho_{0} \sigma_{0}^2}=\frac{\dot M_1 \times v^{\rm wind}_1}{\dot M_0 \times v^{\rm wind}_0}\;.
 \end{equation}
 If equation \ref{eq:scaling1} is satisfied, the two outflows will have the same shape. 
 But the total amount of gas contained in the outflows and their velocity are different.
 The amount of gas is proportional to the entrainment rate and proportional to the outflow age, therefore 
 \begin{equation}\label{eq:scaling2}
 \frac{M_1}{M_0}=\frac{\rho_1 \sigma_1 t_1 \alpha_1}{\rho_0 \sigma_0 t_0
 \alpha_0}\;.
 \end{equation}
 
As discussed in \S \ref{sec:massvelo}, the velocity of the outflow is
proportional to the velocity dispersion of the envelope,
\begin{equation}\label{eq:scaling3}
\frac{v_1}{v_0}=\frac{\sigma_{1}}{\sigma_0}\;.
\end{equation}

These scaling relations help us to put outflows with different masses and velocities  into a common picture.

\section{Conclusions}\label{sec:conclu}
We studied the interaction between the wind from a protostar and
the ambient gas in the form of turbulent mixing and proposed a wind-driven
turbulent entrainment model for protostellar outflows.
In the model, the wind from a protostar is in hydrostatic balance with the gas
in the turbulent envelope, and the outflowing gas is completely contained in the
turbulent entrainment layer that develops between the wind and the envelope.
Turbulence in the ambient gas plays two roles in our model: first, turbulent
motion determines the shape of the outflow (\S \ref{sec:shape}). Second,
turbulence contributes to the mass growth of the mixing layer (\S
\ref{sec:conserve}). Our model is a universal one in that it can explain the
outflows from both low and high-mass protostars (\S \ref{sec:discu}).

Our model can reproduce the geometry and kinematic structure of the observed
outflows (\S \ref{sec:obs}, Figure \ref{fig:image_spec}, \ref{fig:grid_image}
and \ref{fig:grid_pv}).
The main results from this study can be summarized as follows:
\begin{enumerate}
\item At the physical scale of an outflow, the average ram-pressure of wind is
similar to the average ram-pressure of its envelope (\S \ref{sec:shape},
\ref{sec:collimation}).

\item The opening angle of the outflow is dependent on the pressure balance
between the wind and the envelope, therefore it can evolve if the pressure of
the wind or the pressure of the envelope changes  (\S \ref{sec:shape},
\ref{sec:collimation}).

\item The ram-pressure of the wind tends to decollimate the outflow, the
ram-pressure of the environment tends to {\it collimate} the outflow (\S
\ref{sec:shape}, \ref{sec:collimation}).

\item At one given point in the observed image of the outflow, the emission has
a wide spread in velocity (\S \ref{sec:obs}, Figure \ref{fig:grid_pv}). Both
high-velocity gas in the close vicinity of the protostar and low-velocity gas in
regions far from the protostar exist as the results of the turbulent entrainment
process.

\item If the outflow is formed through the turbulent entrainment process
outlined here, the velocity spread of the outflow is expected to be about times
the velocity spread of the ambient gas (\S \ref{sec:massvelo}).
This is independent of the other physical parameters, such as the strength of
the wind and the density of the region.

\item In clustered star-forming regions, we expect a population of
dwarf outflows that are confined by the pressure of their environment into
small regions (\S \ref{sec:drawf}).

\end{enumerate}

The universality of the entrainment process motivates speculation that the same
entrainment process may be at work in AGN-driven outflows
\citep[][]{2011ApJ...735...88A,2012ApJ...752...38T}. We suggest that the
turbulent entrainment process works ubiquitously in nature and plays an
important role in shaping outflows in many different situations.

\begin{acknowledgements} 
We thank the anonymous referee for his/her insightful comments, which helped us to improve our draft.
Guang-Xing Li is supported for this research through a stipend
from the International Max Planck Research School (IMPRS) for Astronomy
and Astrophysics at the Universities of Bonn and Cologne.
\end{acknowledgements}

\begin{appendix}

\section{Effect of centrifugal forces}\label{sec:cf}
As the outflowing gas moves along the cavity, it exerts a centrifugal force
on its surrounding. This effect has been considered in several previous works
\citep[e.g. ]{1980A&A....86..327C,1980ApJ...239..982C,1993RMxAA..25...95B}. The
pressure produced by the centrifugal force can be estimated as
\begin{equation}
p_{\rm centrifugal}\sim v_{\rm outflow}^2 \kappa \rho_{\rm outflow}  d\;,
\end{equation}
where $v_{\rm outflow}$ is the velocity of the outflow, $\kappa$ is the curvature of the outflow cavity, 
$\rho_{\rm outflow}$ is the density of the outflow, and $d$ is the thickness of the outflow. 

The ram-pressure of the envelope is 
\begin{equation}
p_{\rm envelope}\sim \rho_{\rm envelope}\times \sigma_{\rm turb}^2\;.
\end{equation}

In our entrainment model, the density of the outflow is similar to the density
of the envelope $\rho_{\rm outflow} \sim \rho_{\rm envelope}$, and the velocity of the outflow is several times
the velocity dispersion of the envelope (Equation \ref{eq:outflowv}).
Therefore, the ratio between centrifugal pressure and the pressure of the envelope is 

\begin{equation}\label{equation:cf}
f=\frac{p_{\rm centrifugal}}{p_{\rm envelope}}\sim\frac{\rho_{\rm outflow} \kappa d v_{\rm outflow}^2}{\rho_{\rm envelope} \sigma_{\rm turb}^2}\sim 10 \times d \times \kappa \sim 10\times \frac{d}{R}\;,
\end{equation}

where $R$ is the curvature radius. Here we are interested in making
order-of-magnitude estimations, and the effect of projection on the pressure is
neglected. Therefore we have $p_{\rm wind}\sim p_{\rm envelope}$.
In our calculation, $R$ is approximately the size of the outflow and $d$ is
the thickness of the outflow.
Here, we take advantage of the fact that $\rho_{\rm outflow}\sim \rho_{\rm
envelope}$ and $v_{\rm outflow}\sim 3 v_{\rm envelope}$. It can be seen that if
$d\sim R$, centrifugal force will play  an important role in changing the shape
of the outflow cavity.
In our case, since $d \ll R$ (Fig. \ref{fig:density}, note that the thickness of
the outflow has been artificially scaled up for clarity, and in fact our
outflow layer is very thin compared with the size of the outflow), the influence
of centrifugal force is within a few percent and is therefore generally
insignificant.

Our case is different from the case of
\citet{1980ApJ...239..982C} and \citet{1993RMxAA..25...95B} who found the
centrifugal force to be important.
This is because our entrainment process conserves momentum and at the same time
increases the mass of the outflow significantly, and in
\citet{1980ApJ...239..982C} and \citet{1993RMxAA..25...95B} the entrainment
process is insignificant.
To illustrate the effect of mass growth on the centrifugal force, we consider a
particle of mass $m$ rotating along a circle of radius $R$ with velocity $v$.
The centrifugal force can be expressed as
\begin{equation}
f=\frac{m v^2}{R}=\frac{p^2}{m R}\;,
\end{equation}
where $p$ is magnitude of the momentum of the particle.
Here, it is quite clear that when the momentum is conserved, the larger the
mass, the weaker the centrifugal force.

In our case, the entrainment process conserves momentum, but the mass of the
outflow has been increased by a huge factor. Assuming pressure balance $p_{\rm
wind}\sim p_{\rm envelope}$, we have $\rho_{\rm wind} v_{\rm wind}^2 \sim
\rho_{\rm envelope} \sigma_{\rm envelope}^2$.
The mass from the wind is $m_{\rm wind}\sim \rho_{\rm wind} v_{\rm wind} \beta
\times {\rm d} s$, where ${\rm d}s$ represents the surface area, and the mass
from the envelope is $m_{\rm envelope} \sim  \alpha \times \rho_{\rm envelope}
\sigma_{\rm envelope} \time \alpha \times {\rm d} s$. The ratio $m_{\rm
envelope}/m_{\rm wind}$ is
\begin{eqnarray}
\frac{m_{\rm envelope}}{m_{\rm wind}} \sim  \frac{\rho_{\rm envelope}\times v_{\rm envelope}}{\rho_{\rm wind}\times v_{\rm wind}} \\ \nonumber
\sim \frac{p_{\rm envelope}/\sigma_{\rm envelope}\times \alpha}{ p_{\rm wind}/v_{\rm wind}\times \beta} 
\sim   \frac{1}{3}\times \frac{v_{\rm wind}}{\sigma_{\rm envelope}}\;.
\end{eqnarray}
Since $v_{\rm wind}>> \sigma_{\rm envelope}$,
the mass growth from the entrainment process is huge, therefore the centrifugal
force becomes insignificant.

 As the wind gas mixes with the ambient gas, the momentum of the system is
conserved, but the mass of the system becomes much larger, therefore the
centrifugal force becomes much smaller.  This is because the momentum-conserving
entrainment process increases the outflow mass significantly so that the
centrifugal force is insignificant.

\end{appendix}
\bibliographystyle{aa} 
\bibliography{paper}

\end{document}